\documentclass[letterpaper, 10 pt, conference]{ieeeconf}  % Comment this line out if you need a4paper

\IEEEoverridecommandlockouts                              % This command is only needed if 
\usepackage{graphics} % for pdf, bitmapped graphics files
\usepackage{epsfig} % for postscript graphics files
\usepackage{times} % assumes new font selection scheme installed
\usepackage{amsmath} % assumes amsmath package installed
\usepackage{amssymb}  % assumes amsmath package installed
\usepackage{arydshln}
\usepackage{multirow}

\usepackage[nobiblatex]{xurl}

\title{\LARGE \bf
LC-GAN: Image-to-image Translation Based on Generative Adversarial Network for Endoscopic Images
}

\author{Shan Lin, Fangbo Qin, Yangming Li, Randall A. Bly, Kris S. Moe, Blake Hannaford$^*$, \textit{Fellow, IEEE} % <-this % stops a space
\thanks{Shan Lin, Randall A. Bly, Kris S. Moe and Blake Hannaford are with University of Washington, Seattle, WA 98195, USA.
        {\tt\small {shanl3}@uw.edu}}%
\thanks{Fangbo Qin is with Institute of Automation, Chinese Academy of Sciences, Beijing 100190, China
        {\tt\small qinfangbo2013@ia.ac.cn}}%
\thanks{Yangming Li is with Rochester Institute of Technology, Rochester, NY 14623, USA
        {\tt\small Yangming.Li@rit.edu}}%
}

\begin{document}

\maketitle
\thispagestyle{empty}
\pagestyle{empty}

%%%%%%%%%%%%%%%%%%%%%%%%%%%%%%%%%%%%%%%%%%%%%%%%%%%%%%%%%%%%%%%%%%%%%%%%%%%%%%%%
\begin{abstract}

%Visual servoing that uses surgical instrument segmentation and detection to guide the robotic devices for surgery is attractive because this approach is non-invasive and requires minimal modification to the surgical workflow. 
Intelligent vision is appealing in computer-assisted and robotic surgeries. Vision-based analysis with deep learning usually requires large labeled datasets, but manual data labeling is expensive and time-consuming in medical problems. We investigate a novel cross-domain strategy to reduce the need for manual data labeling by proposing an image-to-image translation model live-cadaver GAN (LC-GAN) based on generative adversarial networks (GANs). We consider a situation when a labeled cadaveric surgery dataset is available while the task is instrument segmentation on an unlabeled live surgery dataset. 
We train LC-GAN to learn the mappings between the cadaveric and live images. For live image segmentation, we first translate the live images to fake-cadaveric images with LC-GAN and then perform segmentation on the fake-cadaveric images with models trained on the real cadaveric dataset. The proposed method fully makes use of the labeled cadaveric dataset for live image segmentation without the need to label the live dataset. LC-GAN has two generators with different architectures that leverage the deep feature representation learned from the cadaveric image based segmentation task. Moreover, we propose the structural similarity loss and segmentation consistency loss to improve the semantic consistency during translation. Our model achieves better image-to-image translation and leads to improved segmentation performance in the proposed cross-domain segmentation task.
%and perform instrument segmentation on fake-cadaveric surgery images translated from the live surgery images for quantitative analysis. 
%The segmentation models are trained with the cadaveric dataset so our strategy does not require labeling of the live dataset. 
%to leverage the deep feature representations learned from the labeled cadaveric dataset.

%leads to better segmentation performance than the other three variants. 
%When applying a trained model to a different but relevant dataset, a new labeled dataset may be required for training to prevent the performance from decreasing. 
%While previous relevant works have studied image-to-image translation problems between different imaging modalities and between virtual and real images, ours is the first to study the challenging translation task between the live and cadaveric surgery images.
\end{abstract}

%%%%%%%%%%%%%%%%%%%%%%%%%%%%%%%%%%%%%%%%%%%%%%%%%%%%%%%%%%%%%%%%%%%%%%%%%%%%%%%%
\section{Introduction}
% Segmentation of surgery instrument (bouget review paper)
Minimally invasive surgery (MIS) such as endoscopic surgery brings many benefits to patients, 
%such as minimal access incisions, less trauma, pain and bleeding, as well as a shorter hospital stay \cite{bouget2017vision,mack2001minimally}. However, minimally invasive procedures 
but also introduces many challenges to the surgeons. 
%and the clinical team. 
In endoscopy, a tiny camera is inserted into the human body with surgical instruments to provide a real-time view of the surgical site. The endoscope provides limited field-of-view and reduced depth perception that impacts the surgeons' eye-hand coordination ability \cite{bogdanova2016depth,bouget2017vision}. Additionally, limited sense of touch further reduce the information a surgeon can get from the surgical site \cite{su2019multicamera,rosen1999force}. 

The challenges of MIS may lead to accidental damage to important structures and suboptimal surgical results \cite{runciman2019soft,khanna2019managing}. To improve surgical performance, robotic-assisted surgery systems have attracted increasing attention \cite{friedrich2017recent}. For example, robots that handle the endoscope or laparoscope have been explored \cite{sahu2016instrument,capolei2017positioning}, soft robotic devices have been studied to reduce patients' pain and damage in MIS \cite{runciman2019soft}. Moreover, overlaying pre- and intra-operative imaging with surgical videos could enhance surgeons' capabilities \cite{allan20202018}. For all these applications, instrument segmentation is a critical component and one of the main challenges \cite{bouget2017vision}. 

In recent years, deep learning has achieved cutting-edge performance on instrument segmentation over traditional approaches. 
% based on hand-craft features. 
However, most deep learning-based methods rely on large-scale labeled datasets, whose the availability is usually limited for medical applications \cite{cheplygina2019not,bouget2017vision}. 
When a different but related labeled dataset is available, we explore a less expensive cross-domain strategy that uses domain adaptation to transfer the knowledge learned by a particular model on a source domain to a related target domain \cite{hoffman2018cycada}. Recently, domain adaptation is usually implemented by searching for a common feature space of both domains but this strategy suffers from semantic inconsistency, i.e. one type of object may be mapped to another type of object \cite{hoffman2018cycada}.

%We will show later in this paper that the segmentation performance on cadaveric surgery images is better than the performance on live surgery images using a deep convolutional neural network DeepLabV3+ \cite{chen2018encoder}. 
In this work, we explore the instrument segmentation task on an unlabeled live surgery dataset using a labeled cadaveric surgery dataset. We propose an image-to-image translation model LC-GAN that learns the mapping between the cadaveric and live image domains. %We quantitatively evaluated our method through instrument segmentation. 
Our model is developed based on CycleGAN \cite{zhu2017unpaired} that learns the bijective translations between two image domains.
%by Zhu \emph{et al.}
%
%First, we construct two generators with different architectures for LC-GAN. Based on our assumption that only the labels of the source domain are available, we provide better feature representations using Resnet-50 \cite{he2016identity} pre-trained on the source domain to the generator that translates images from source domain to target domain. For the other generator, we choose a Resnet architecture and train it from scratch. 
First, we improve the feature representation of one generator using a backbone feature extractor trained on the cadaveric domain.
Secondly, we propose a structural similarity loss to reduce image structural changes in translation and a segmentation consistency loss to encourage semantic consistency between the real and the generated fake images. 
%These two loss functions together improve the semantic consistency of LC-GAN. 
%
We then use the deep segmentation models trained on the cadaveric domain to segment the fake-cadaveric surgery images translated from the live domain. 
%combined with a multi-angle feature aggregation (MAFA) strategy \cite{2020arXiv200210675Q}
%
We demonstrate our method on a novel sinus surgery dataset.
%with 10 cadaveric surgery videos as the source domain and 3 live surgery videos as the target domain. 

To the best of our knowledge, our work is the first to perform image-to-image translation between live and cadaveric surgery images. In addition to the proposed cross-domain strategy, LC-GAN could be used in data augmentation or to generate a larger dataset by combining two smaller ones. Also, our work could bring benefits to the surgery training processes. For example, we could augment the cadaveric surgery videos in real-time to live surgery-like videos as the surgical trainees gain experience on the cadaver. 

\section{Related Works}

\textbf{Generative adversarial networks (GANs)} proposed by Goodfellow \emph{et al.} \cite{goodfellow2014generative} have achieved impressive performance in many vision problems such as image generation, feature learning and image super-resolution \cite{xu2018attngan,qin2019surgical,ledig2017photo}. GANs train two networks, a generator and a discriminator, which contest with each other. The discriminator is trained to distinguish the generated data from the true data, while the generator is trained to fool the discriminator. 

\textbf{Image-to-Image Translation}
%Image-to-image translation is a vision problem that 
aims to learn the mapping between two relevant image domains \cite{liu2017unsupervised}. GANs have been successfully applied in various image-to-image translation tasks with paired or unpaired datasets \cite{isola2017image,zhu2017unpaired,liu2017unsupervised}. 
%CycleGAN is highly attractive because it using unpaired image data. 

The availability of certain image modalities may be limited in medical applications because data collection is usually expensive and time-consuming \cite{cheplygina2019not,bouget2017vision}. Image-to-image translation provides an effective way to estimate the appearance of the desired image modality from a relevant image and has brought impressive results in medical image analysis problems \cite{wang2018conditional,mahmood2018unsupervised,li2018cc}. 
%funke2018generative
Wang \emph{et al.} used conditional GAN to reduce the artifacts from computed tomography (CT) and achieved better segmentation results on the generated CTs \cite{wang2018conditional}. Mahmood \emph{et al.} applied GAN to transform true medical images to synthetic-like images and estimated the depth maps using a model trained on synthetic images \cite{mahmood2018unsupervised}. 

The unreliability of the generated images due to semantic inconsistency in translation leads to the concern of applying image-to-image translation to medicine. Recent studies worked on introducing constraints to guarantee semantic consistency. In CycleGAN \cite{zhu2017unpaired}, a cycle consistency loss was introduced to ensure the translation is invertible. Liu \emph{et al.} included an assumption that the two image domains share a latent space based on CycleGAN \cite{liu2017unsupervised}. Cherian \emph{et al.} used two segmentation models alongside the CycleGAN to enhance semantic consistency in the training paradigm \cite{cherian2019sem}.

\textbf{Surgical Instrument Segmentation} %Vision-based surgical instrument segmentation has been widely studied and 
%The traditional machine learning methods such as random forest (RF) have been used based on hand-crafted features include color, texture and gradient information \cite{bodenstedt2018comparative,bouget2017vision,bouget2015detecting}.
Deep learning models have achieved cutting-edge instrument segmentation performance \cite{long2015fully,ronneberger2015u,garcia2017toolnet,chaurasia2017linknet,iglovikov2018ternausnet,chen2018encoder,shvets2018automatic,islam2019learning,allan20192017}. 
%Since Fully Convolutional Networks (FCN) \cite{long2015fully} and then immediately U-Net \cite{ronneberger2015u} were proposed for image segmentation, many models such as Toolnet \cite{garcia2017toolnet}, LinkNet \cite{chaurasia2017linknet} and Ternausnet \cite{iglovikov2018ternausnet} were constructed based on these two structures. 
Shvets \emph{et al.} improved two U-Net \cite{ronneberger2015u} family models Ternausnet and LinkNet and achieved top performances in the 2017 MICCAI Sub-Challenge on Robotic Instrument Segmentation \cite{shvets2018automatic,allan20192017}. Chen \emph{et al.} proposed an advanced segmentation model DeepLabV3+ using the pyramid pooling module \cite{zhao2017pyramid} and atrous convolution \cite{chen2018encoder}. 
%by Zhao \emph{et al.}
Despite these efforts, existing algorithms may not be robust enough under challenging conditions in surgeries such as strong specular reflection and blood. Additionally, the trained models may not generalize well to different surgical interventions. To achieve good performance, labeling a new and large dataset is needed, which requires significant expert time and can be very expensive.

\section{Methods}

In this work, we explore image-to-image translation to reduce the need for manual data labeling for surgical instrument segmentation. We consider a situation that there exists a labeled dataset and one wants to perform segmentation on another related but different unlabeled dataset. The two datasets used here are cadaveric surgery images and live surgery images. Although we have manually labeled both datasets, we assume the labels of the cadaveric dataset are available while the labels of the live dataset are only used to evaluate the proposed method. Section \ref{dataset} provides more details on the datasets. 

\begin{figure*}
\centering
\includegraphics[width=0.99\textwidth]{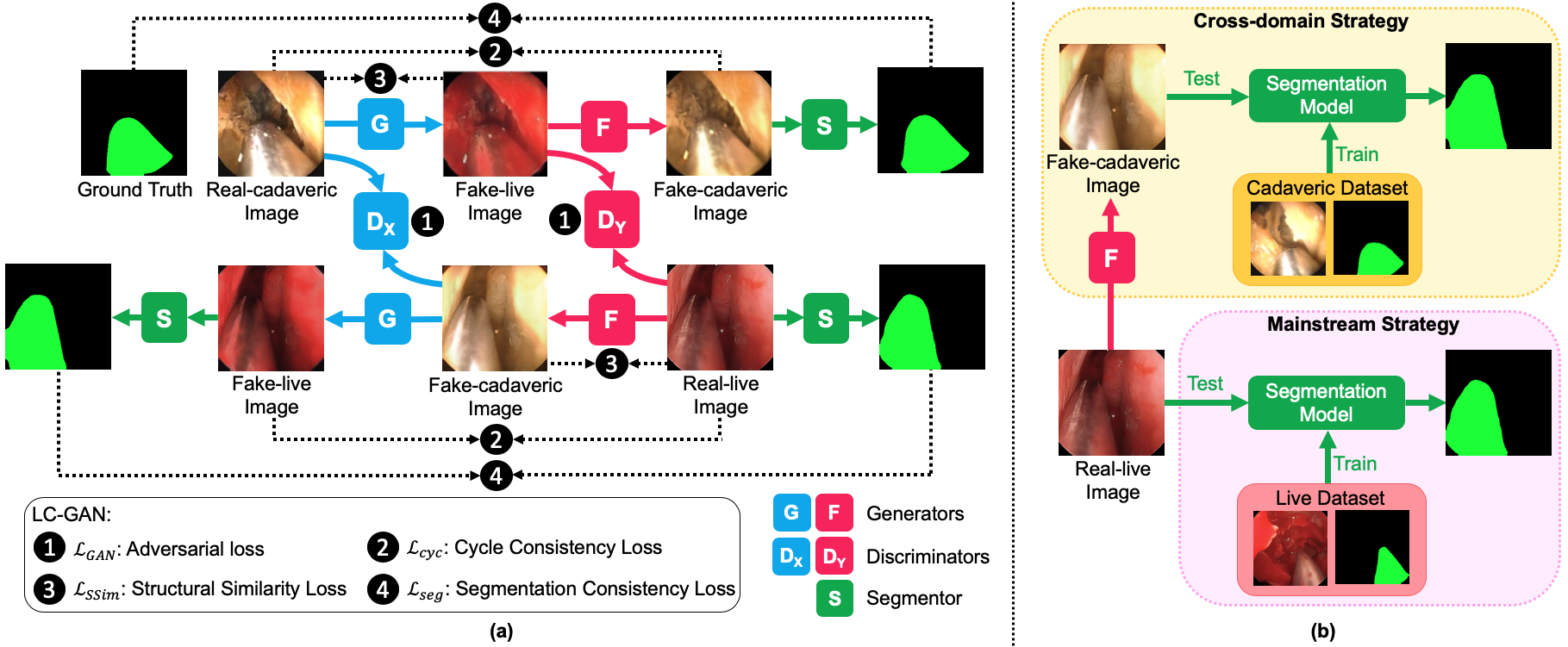}
\vspace{-1.em}
\caption{Overall framework. (a) Schematic of LC-GAN. The generator $G$ performs cadaver-to-live translation, while generator $F$ performs live-to-cadaver translation. The discriminators $D_X$ and $D_Y$ are used to distinguish the fake images from the real images in cadaveric and live domains, respectively. The segmentor S is a deep segmentation model trained on the real cadaveric dataset. (b) Schematic of the proposed cross-domain strategy and the mainstream strategy for instrument segmentation on the live dataset. The predicted or ground truth instrument regions in the segmentation masks are shown in green. Note that there is only one instance of G, F, and S, respectively. They are repeated to avoid cross-connection.} 
\vspace{-0.5em}
\label{workflow}
\end{figure*}

The overall framework is shown in Fig. \ref{workflow}. The task is instrument segmentation on the live surgery images. As shown in Fig. \ref{workflow}(b), the current mainstream strategy is to train and test segmentation models directly on a labeled live dataset, but this strategy requires us to also label the live dataset for good performance. In contrast, we propose a cross-domain strategy that does not use the labels of live dataset by taking advantage of the already existed labeled cadaveric dataset as shown in Fig. \ref{workflow}(b). We propose a model LC-GAN (see Fig. \ref{workflow}(a)) developed based on CycleGAN \cite{zhu2017unpaired} to learn the mapping between the cadaveric dataset and live dataset. We then perform instrument segmentation on the fake-cadaveric surgery images generated from the real-live surgery images. The segmentation is implemented using deep convolutional neural network (CNN) models trained with the labeled cadaveric dataset. Finally, we evaluate the potential of the cross-domain strategy by comparing its segmentation results with the current mainstream strategy.

%The target task is to perform instrument segmentation on the live dataset. We proposed a strategy that doesn't require the labels of live dataset. We first implement image-to-image translation with a model Asym-CycleGAN to generate fake-cadaveric surgery images from real-live surgery images. We then use the segmentation models trained on the already existed labeled cadaveric dataset to segment the fake images. We compare our method with the traditional strategy that trains and tests on the labeled live dataset.

\subsection{Live-cadaver GAN (LC-GAN) Architecture}

\begin{figure}
\centering
\includegraphics[width=0.48\textwidth]{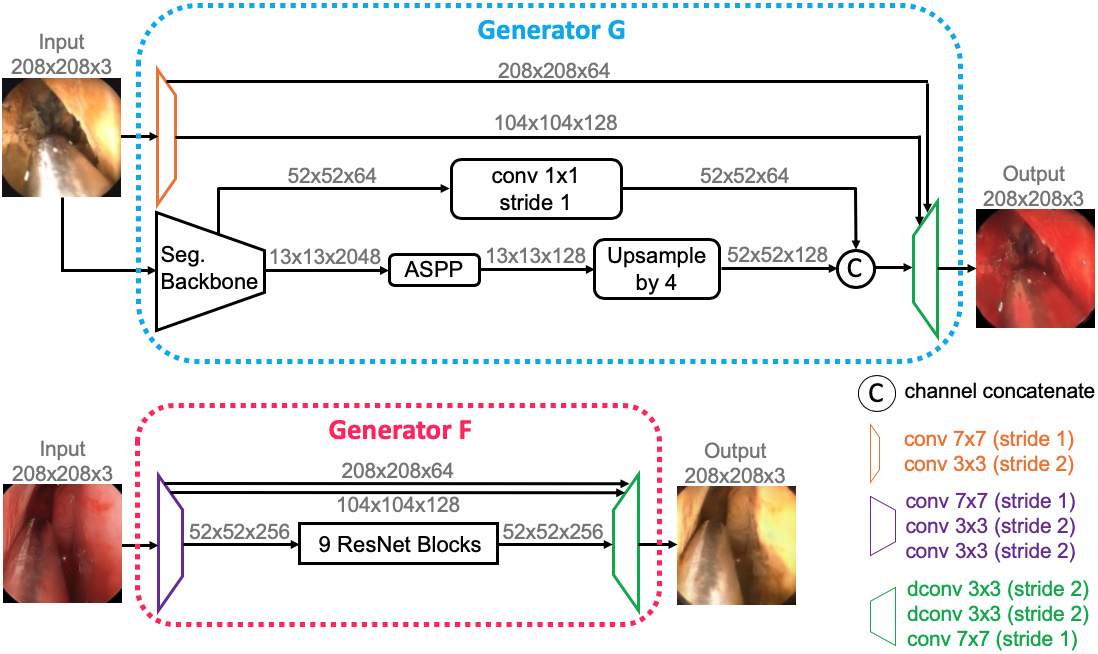}
\vspace{-2.em}
\caption{Generator architectures of LC-GAN (example for 208x208 input image). Each trapezoid represents a series of convolution or deconvolution operations. The sizes (width-height-channel) of the feature maps are shown on top of the corresponding arrows.} 
\vspace{-1.5em}
\label{gen_struc}
\end{figure}

Similar to CycleGAN \cite{zhu2017unpaired}, LC-GAN 
%We developed an image-to-image translation model based on the framework of CycleGAN proposed by Zhu \emph{et al.} \cite{zhu2017unpaired}. 
consists of two generators and two discriminators. Given two image domains $X$ and $Y$, the two generators $G$ and $F$ are trained to learn the mappings between the image domains, i.e. $G: X\rightarrow Y$ and $F: Y\rightarrow X$. $D_X$ and $D_Y$ are two discriminators trained to distinguish between real and fake images. We define the cadaveric dataset as the $X$ domain and the live dataset as the $Y$ domain.

For images with many challenging conditions as in our dataset, training the generators from scratch is difficult and may require longer training time to converge. Considering that we already have deep segmentation models trained on the cadaveric dataset, we use the backbone feature extractors in these models to provide better feature maps for the generator. 
%Unlike CycleGAN \cite{zhu2017unpaired}, 
%Inspired by \cite{dou2019asymmetric}, 
We construct two generators with different architectures as shown in Fig. \ref{gen_struc}. 
For generator $G$ that maps cadaveric surgery images to live surgery images, we use the backbone of the segmentation model trained with the labeled cadaveric dataset as the feature extractor. 
%For the generator $G$ that maps cadaveric surgery images to live surgery images, we use the network backbone Resnet-50 \cite{he2016identity} of a DeepLabV3+ segmentation model \cite{chen2018encoder} trained with the labeled cadaveric dataset as the feature extractor.
We also adapt part of the DeepLabV3+ segmentation model \cite{chen2018encoder} into $G$. We use the Atrous Spatial Pyramid Pooling (ASPP) module \cite{chen2018encoder} that extracts multi-scale information from the output of Resnet-50, and concatenate this information with low-level features from Resnet-50. The parameters of Resnet-50 are fixed during training.
For another generator $F$, no trained feature extractor is available so we train it from scratch. We choose a ResNet with two stride-2 convolutions, nine residual blocks and two fractionally-strided convolutions with stride $\frac{1}{2}$ as the generator $F$ \cite{he2016deep,zhu2017unpaired}. 
For the discriminator, we use the 70$\times$70 PatchGAN \cite{isola2017image,zhu2017unpaired} to determine the real 70$\times$70 image patches from fake patches.

%The adversarial loss is the objective function of GANs \cite{goodfellow2014generative}. For the mapping from X to Y, the adversarial loss is expressed as
% \begin{equation}
% \begin{align}
% \mathcal{L}_{GAN}(G,D_Y) &= \mathbb{E}_{y\sim Y}[\log D_Y(y)] +\\
% & \mathbb{E}_{x\sim X}[\log(1 - D_Y(G(X)))]
% \end{align}
% \end{equation}
% One contribution of CycleGAN is the cycle consistency loss that improves semantic consistency in translation (see Fig. \ref{loss}(a)). The cycle consistency loss is based on the expectation that if an image is translated to another domain and then translated back, the output image should be similar to the input. The cycle consistency loss can be expressed as
% \begin{equation}
% \begin{align}
% \mathcal{L}_{cyc}(G,F) &= \mathbb{E}_{x\sim X}[\|F(G(x))-x\|_1] +\\
% & \mathbb{E}_{y\sim Y}[\|G(F(y))-y\|_1]
% \end{align}
% \end{equation}

% [? reveals the difficulty of GAN in medical area: the liability of image is not confirmed. tissues could be mapped into tool and vise versa. The cycle consistency loss in CycleGAN doesn't necessary the object to remain same semantic information during translation. Therefore, more constraints are required to improve the quality of generated fake images. the xxx function in the proposed method surved as for this goal. ]

\subsection{Loss Functions for LC-GAN}

%Examples of semantic inconsistency in image-to-image translation using CycleGAN. The bone region pointed by the white arrow is translated into an instrument and the instrument becomes much larger in the fake image than its true size.

\begin{figure}
\centering
\includegraphics[width=0.4\textwidth]{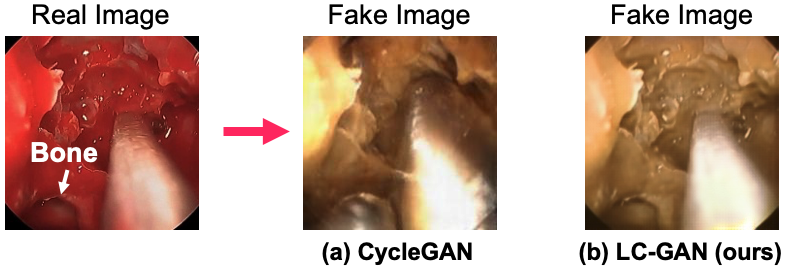}
\vspace{-1.em}
\caption{Translation from a real-live surgery image to a fake-cadaveric surgery image. (a) The result of CycleGAN is an example of semantic inconsistency. The bone region pointed by the white arrow is translated into an instrument and the instrument becomes much larger in the fake image than its true size. (b) The proposed LC-GAN generates a fake image with better semantic consistency.}
\vspace{-1em}
\label{semantic}
\end{figure}

%Multi-scale Structural Dissimilarity Loss and Segmentation Loss
CycleGAN was proposed with the adversarial loss $\mathcal{L}_{GAN}$ and the cycle consistency loss $\mathcal{L}_{cyc}$. The details of $\mathcal{L}_{GAN}$ and $\mathcal{L}_{cyc}$ are provided in \cite{zhu2017unpaired}. Although CycleGAN shows compelling results in many datasets, the semantic consistency might not be guaranteed for some complex scenes. Fig. \ref{semantic}(a) shows an example of semantic inconsistency between the real-live surgery image and the fake-cadaveric surgery image generated using CycleGAN. 
%In Fig. \ref{semantic}(a), the instrument is translated into an area similar to the background. 
In Fig. \ref{semantic}(a), the bone is translated to an instrument and the true instrument becomes much larger in the fake image.
%
%The reason for these problems is that both generators learn the incorrect and reciprocal mappings. For example, in Fig. \ref{semantic}, one generator learns a mapping of 'instrument $\rightarrow$ background', while another generator learns a mapping of 'background $\rightarrow$ instrument'.
To ensure the semantic consistency, we propose structural similarity loss $\mathcal{L}_{SSim}$ and segmentation consistency loss $\mathcal{L}_{seg}$ alongside with the cycle consistency loss as shown in Fig. \ref{workflow}(a). 
%shows the schematic of the loss functions for LC-GAN. 

$\mathcal{L}_{SSim}$ estimates the structural similarity between the input and output of a generator. Because the live and cadaveric surgery images have very different colors, the color information should be excluded when calculating $\mathcal{L}_{SSim}$. 
%Considering in the YUV color space, the Y channel consists of luminance information and the other two channels are chrominance components \cite{ford1998colour}. 
Therefore, we convert the image from RGB to YUV color space and use the Y channel that consists of luminance information \cite{ford1998colour} to estimate the structural similarity. Besides color features, the live surgery images have different lighting conditions and the presence of fluids including blood 
%present a more moisture environment 
%with fluids such as blood 
compared to the cadaveric surgery images. Therefore, $\mathcal{L}_{SSim}$ should focus on overall structural similarity while allowing differences in image brightness, contrast and details. Similar to \cite{wang2003multiscale,pfeiffer2019generating}, we propose a multi-scale method that compares the image structural information at different resolutions. We iteratively downscale the image by a factor of 2 and get a total of $n_s$ scaled images which are 1, $\frac{1}{2}$, ..., $\frac{1}{2^{n_s}}$ of the original size, respectively. 
%For the mapping from $X$ to $Y$, 
The expression of $\mathcal{L}_{SSim}$ is
\begin{equation}
\begin{aligned}
\mathcal{L}_{SSim}(G,F) &= [1-\sum_{i=0}^{n_s}\gamma_iC(x_{Y,i},G(x)_{Y,i})]\\
&+[1-\sum_{i=0}^{n_s}\gamma_iC(y_{Y,i},F(y)_{Y,i})]
\end{aligned}
\end{equation}
where $x\in X$ is an image from the $X$ domain and $y\in Y$ is an image from the $Y$ domain, the subscripts $Y,i$ of an image denote that we extract the Y channel of the image and scale it to $\frac{1}{2^{i}}$ of its original size, $\gamma_i$ are the multi-scale weights and are normalized to $\sum_{i=0}^N\gamma_i=1$.
%This structural dissimilarity loss $\mathcal{L}_{DSSIM}$ is added to the total loss function with a weight of $\lambda_2$. 
$C(a,b)$ is the zero-normalized cross-correlation (ZNCC) \cite{giachetti2000matching,wang2004image} between images $a$ and $b$
\begin{equation}
C(a,b) = \frac{\sigma_{ab}+\epsilon}{\sigma_a\sigma_b+\epsilon}
\end{equation}
where $\sigma_{ab}$ is the covariance between $a$ and $b$, and is defined as \cite{wang2004image}
\begin{equation}
\sigma_{ab} = \frac{1}{m-1}\sum_{i=1}^{m}(a_i-\mu_a)(b_i-\mu_b)
\end{equation}
$m$ is the number of pixels in $a$ and $b$, $a_i$ and $b_i$ are the $i$th pixel of $a$ and $b$. $\mu_a$, $\mu_b$ and $\sigma_a$, $\sigma_b$ correspond to the mean intensity and standard deviations of $a$ and $b$. $\epsilon$ is a constant to stabilize the division when $\sigma_a\sigma_b$ is close to zero. ZNCC is less sensitive to the differences of illumination conditions and contrast of the two compared images \cite{wang2004image}. 
%
%Moreover, many images in our cadaveric dataset have a large white area due to overexposure, while such large white areas do not exist in the live dataset. Therefore, the large white areas in cadaveric surgery images were not considered when calculate $\mathcal{L}_{SSim}$.

% \begin{figure}
% \centering
% \includegraphics[width=0.35\textwidth]{./plots/losses.png}
% \caption{The schematic of the loss functions for Asym-CycleGAN. We only show the loss functions involve in the path of real-cadaveric $\rightarrow$ fake-live $\rightarrow$ fake-cadaveric. We use the same loss functions for path real-live $\rightarrow$ fake-cadaveric $\rightarrow$ fake-live. (a) Cycle consistency loss. (b) Structural dissimilarity loss. (c) Segmentation loss. (d) Adversarial loss.} \label{loss}
% \end{figure}

Inspired by \cite{cherian2019sem,ramirez2018exploiting}, we introduce a segmentation consistency loss to further improve semantic consistency:
\begin{equation}
\begin{aligned}
\mathcal{L}_{seg}(G,F) &= \mathcal{L}_{CE}( x_m,S(F(G(x))) )\\
& +\mathcal{L}_{CE}( S(y),S(G(F(y))) )
\end{aligned}
\end{equation}
where $x_m$ is the ground truth segmentation mask of image $x$, $S$ is a segmentation model trained on the labeled cadaveric dataset, $\mathcal{L}_{CE}(a,b)$ is the naive cross-entropy loss. The parameters of $S$ are fixed during LC-GAN training. For the segmentation consistency loss between $y$ and $G(F(y))$, we use the segmentation on the image $y$ as the target mask because the labels of the live dataset are not available according to our assumption. 
%instead of using the ground truth segmentation mask of image $y$.

Finally, the overall objective function is given by 
\begin{equation}
\begin{aligned}
\mathcal{L}(G,F,D_X,D_Y)&= \mathcal{L}_{GAN}(G,D_Y)+\mathcal{L}_{GAN}(F,D_X)\\
& +\lambda_1 \mathcal{L}_{cyc}(G,F)+\lambda_2\mathcal{L}_{SSim}(G,F)\\
&+\lambda_3\mathcal{L}_{seg}(G,F)\\
\end{aligned}
\end{equation}
where $\lambda_i$ are hyper-parameters that balance the impact of the losses. The generators are trained to minimize the overall objective function and the discriminators are trained to maximize it.

\section{Datasets}\label{dataset}

\begin{figure}
\centering
\includegraphics[width=0.5\textwidth]{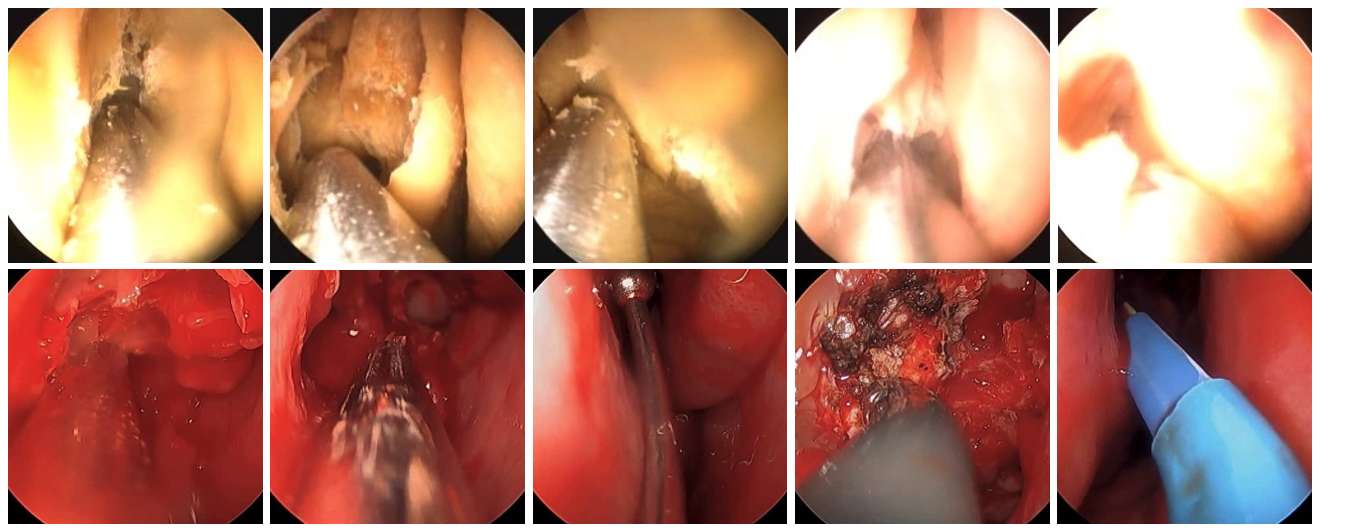}
\vspace{-1.em}
\caption{Examples of center-cropped video frames in our dataset. The top row is from the cadaveric surgery videos and the bottom row is from the live surgery videos.} 
\vspace{-1.6em}
\label{exs}
\end{figure}

Our sinus surgery dataset consists of 10 cadaveric surgery videos ranging from 5 minutes to 23 minutes with a resolution of 320$\times$240 and 3 live surgery videos ranged from 12 minutes to 66 minutes with a resolution of 1920$\times$1080. More details of data collection are provided in the author's previous publication \cite{lin2019automatic}. Fig. \ref{exs} shows some examples of video frames. The challenging conditions of this dataset include specular reflections, blur from motion, blood, smoke, instruments in shadow and occlusions by tissues. 
%Bone, titanium mesh, and gauze in the background make the segmentation on live surgery images very difficult. Fig. \ref{diff_exs} shows some examples of these challenging conditions. 

The video frames were sampled from the cadaveric surgery videos at 0.5 Hz resulting in a total of 4345 frames. For the live surgery videos, due to blood, blur from motion and specular reflection, the instrument regions in some video frames cannot be identified without temporal information from neighboring frames. The temporal information has not been considered in either the image-to-image translation or segmentation methods in this paper, so such images were excluded. %from our experiments. 
We captured frames from the live surgery videos at 1 Hz and removed the aforementioned challenging images resulting in a total of 4658 frames. 

For surgical instrument segmentation, we center-cropped the frames and downscaled them to 240$\times$240. The cadaveric surgery images were separated into a training set of 2760 frames from 7 videos and a test set of 1585 frames from the remaining 3 videos. The live surgery images were separated into a training set of 3504 frames from 2 videos and a test set of 1154 frames from the remaining 1 video. The instrument contours in all frames were manually labeled.

\begin{figure*}[h]
\centering
\includegraphics[width=1\textwidth]{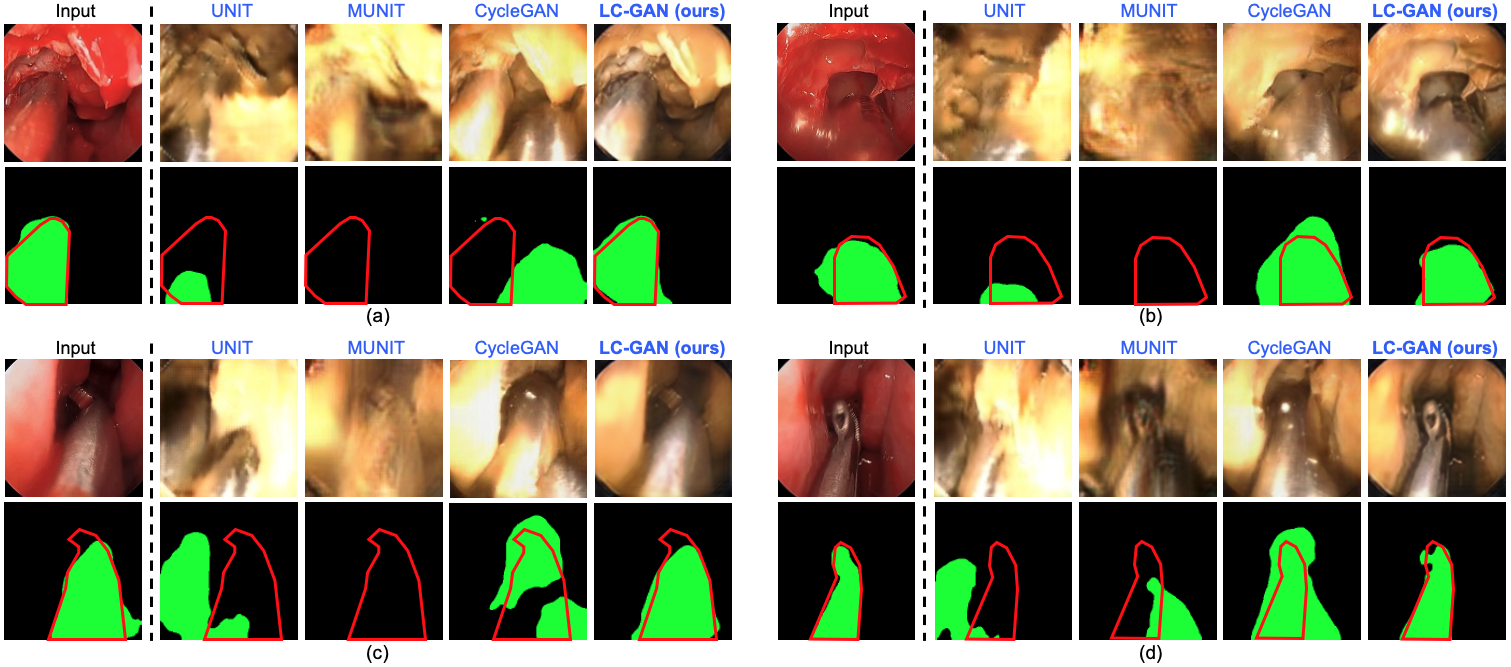}
\vspace{-1.8em}
\caption{Examples of results from the cross-domain strategy and mainstream strategy. In each subfigure, the last four columns of the top row show the fake-cadaveric images translated from the input real-live image using UNIT \cite{liu2017unsupervised}, MUNIT \cite{huang2018multimodal}, CycleGAN \cite{zhu2017unpaired} and LC-GAN (ours). The second row shows the corresponding instrument segmentation results obtained using DeepLabV3+ \cite{chen2018encoder} with MAFA \cite{2020arXiv200210675Q}. In the bottom row, the first segmentation is the result of the mainstream strategy and the last four segmentations are from the cross-domain strategy. The predicted instrument regions are shown in green and the ground truth of the instrument contours are shown as red lines.}
% \vspace{0.2em}
\label{trans_rst_ex}
\end{figure*}

\begin{figure*}[h]
\centering
\includegraphics[width=0.9\textwidth]{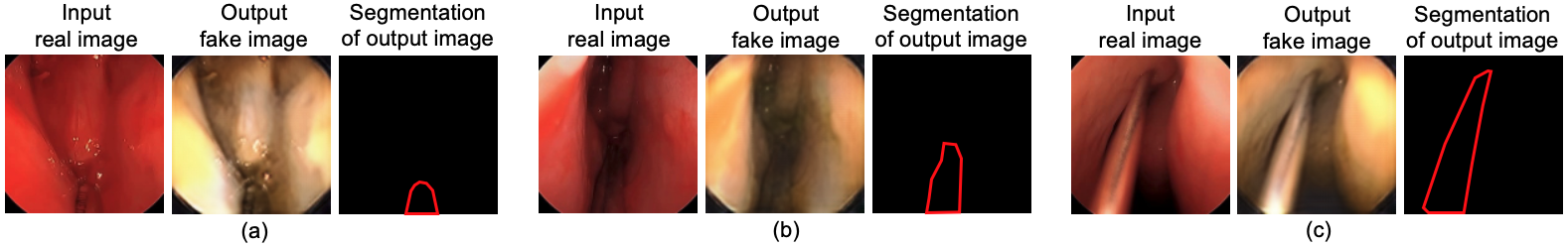}
\vspace{-0.6em}
\caption{Failed live-to-cadaver translation examples given by LC-GAN. The corresponding instrument segmentations are obtained using DeepLabV3+ \cite{chen2018encoder} with MAFA \cite{2020arXiv200210675Q}. The predicted instrument regions are shown in green and the ground truth of the instrument contours are shown as red lines.} 
% \vspace{-0.5em}
\label{fail_ex}
\end{figure*}

Endoscopic images have a large black border region without useful information. To speed up the training process for LC-GAN, we cropped image patches within the endoscopic area and downscaled these patches to 208$\times$208. For training, we selected frames with less specular reflections and blurriness from both cadaveric and live datasets. 900 cadaveric surgery frames were selected for the X domain and 3174 live surgery frames were selected for the Y domain. Note that frames in the live test set for instrument segmentation were excluded in LC-GAN training.

The dataset is available at \url{https://github.com/SURA23/Sinus-Surgery-Endoscopic-Image-Datasets}.

% \begin{figure}
% \centering
% \includegraphics[width=0.5\textwidth]{./plots/diff_exs.png}
% \caption{Challenging conditions: The top row from left to right are blood, smoke, tools in the shadow and occlusions by tissues; the bottom row from left to right are blur from motion, bone, Titanium mesh and gauze in the background.} \label{diff_exs}
% \end{figure}

\section{Implementation Details} \label{implement}
%%%%% training details
We implemented image-to-image translation models using Tensorflow on a single Nvidia Tesla T4 GPU. The segmentation models were implemented on a 3.70GHz Intel
i7-8700K CPU and two Nvidia GTX2080ti GPUs.

% \begin{figure*}[h]
% \centering
% \includegraphics[width=1\textwidth]{./plots/seg_compare.png}
% \caption{Examples of segmentation results of fake-cadaveric images using four segmentation methods DeepLabV3+ \cite{chen2018encoder}, TernausNet \cite{iglovikov2018ternausnet}, MFF+SPP \cite{islam2019real} and LWANet \cite{ni2019attention}. All segmentation methods were tested both with and without MAFA \cite{2020arXiv200210675Q}. } \label{trans_rst_ex}
% \end{figure*}

\textbf{LC-GAN} 
%For data augmentation, we randomly flipped, rotated, zoomed and cropped the training images and expanded the dataset to ten times. 
We trained the network with 6 epochs through Adam optimizer using an initial learning rate of 0.00008 for the first half of training, and then linearly decaying to zero over the remaining epochs. %See Section \ref{dataset} for details of training and test dataset. 
Each batch consisted of one image from domain X and one image from domain Y. 
The hyper-parameters $\lambda_1$, $\lambda_2$ and $\lambda_3$ were set to 5, 1 and 2, respectively. 
% The weight of the cycle consistency loss $\mathcal{L}_{cyc}$, structural dissimilarity loss of mapping from domain X $\mathcal{L}_{DSSIM}(x_Y)$ and the structural dissimilarity loss of mapping from domain Y $\mathcal{L}_{DSSIM}(y_Y)$ are 10, 1 and 0.5, respectively.
For the structural similarity loss $\mathcal{L}_{SSim}$, we chose $n_s=4$ and empirically set $\gamma_1=0.05$, $\gamma_2 = 0.33$, $\gamma_3=0.35$ and $\gamma_4 = 0.27$. 

%To quantitatively evaluate the quality of synthetic data, we make a GUI which randomly shows two real cadaveric surgery images and one synthetic images each time, and let ? surgeons to choose which one is the most different image from the others (see Figure \ref{}). 

% \textbf{UNsupervised Image-to-image Translation Networks (UNIT)}
% We used the same network structure as described in \cite{liu2017unsupervised}. We trained UNIT using Adam optimizer for 10 epochs and a learning rate of 0.00005. We tested UNIT with longer training time until 200 epochs but the results showed that the model became unstable as we increased the epoch number. In training, each batch consists of one image from domain X and one image from domain Y. The weight of the conditional GAN loss, the cycle consistency loss, the adversarial loss and the latent space distribution loss were 10, 100, 10 and 0.1, respectively. 

\textbf{Compared image-to-image translation models}
We compared LC-GAN with three image-to-image translation models CycleGAN \cite{zhu2017unpaired}, UNIT \cite{liu2017unsupervised} and MUNIT \cite{huang2018multimodal}. We first trained all three models until 250 epochs and chose the minimum epochs that made the models converge. We found CycleGAN stabilized at 200 epochs. UNIT and MUNIT failed to converge within 250 epochs. We then chose 9 epochs for these two models because after 9 epochs they became more unstable. 
%used the translation results provided by these two models trained with 9 epochs, because after 9 epochs the models became more unstable. 

%\textbf{DeepLabv3+ with MAFA}
%We trained the model on the cadaveric surgery images for 40 epochs and a batch size of 16 through Adam optimizer based on cross-entropy loss. We used Adam optimizer with an initial learning rate of 0.0005. The learning rate exponentially decayed by a factor of 0.5 for every 10 epochs. We adopted ResNet-50 pretrained on the ImageNet as the encoder. Data augmentation, includes randomly hue, brightness, saturation, contrast changes, horizontal and vertical flipping, cropping and padding, were applied for training. MAFA was implemented with 4 different rotation angles. 
%As a comparison, we also trained a DeepLabV3+ with MAFA model on the live surgery images. The initial learning rate was 0.0001 and other hyperparameters were the same as the training on the cadaveric surgery images. 

%{DeepLabv3+ with Multi-angle Feature Aggregation (MAFA) for Instrument Segmentation}

\section{Experiments and Results}

We compared our proposed method LC-GAN against the baseline and state-of-the-art models include CycleGAN \cite{zhu2017unpaired}, UNIT \cite{liu2017unsupervised} and MUNIT \cite{huang2018multimodal}. 

\textbf{Qualitative evaluation} Fig. \ref{trans_rst_ex} shows examples of the image-to-image translation results. 
%Fig. \ref{fail_ex} shows some failed translation examples generated by the proposed method.
%The results show that DSS-CycleGAN outperforms other image-to-image translation models in this task both qualitatively and quantitatively. 
We found that LC-GAN generally retained the shape and location of the instruments and the structures of sinus tissue in the resultant fake images. A majority of images in our live dataset have the instruments merged into the background due to strong specular reflection and blood. LC-GAN could successfully handle such cases most of the time as shown in Fig. \ref{trans_rst_ex}(a,b). In contrast, CycleGAN was less robust and might translate part of the background to an instrument as shown in Fig. \ref{trans_rst_ex}(a,c). Also, CycleGAN tended to increase the size of instruments as shown in Fig. \ref{trans_rst_ex}(b,d). The other two comparison methods, UNIT and MUNIT, did not converge to the correct correspondence between the two image domains. 

% fail cases
Although our image-to-image translation results are promising, the performance is still not satisfactory under challenging conditions. Fig. \ref{fail_ex} shows some typical fail cases.
%
%Strong specular reflection and blood usually cause a reddish look of the tools. This makes the image extremely challenging when the tools are covered in the shadow. 
Fig. \ref{fail_ex}(a) shows an example of an instrument merged in blood. Fig. \ref{fail_ex}(b) shows an example of an instrument in shadow with a red appearance due to specular reflection. The instruments in Fig. \ref{fail_ex}(a,b) are translated to regions similar to the surrounding background and lead to incorrect segmentations. Fig. \ref{fail_ex}(c) represents another type of failure that happens when the instrument appears red and does not exist in the cadaveric dataset. In such a case, the color information is not reliable and the segmentation models should use shape information for successful segmentation. The segmentation models trained with the cadaveric dataset have not seen this or similar instrument before, resulting in segmentation failure.
%and it is hard for the segmentation model to identify it.

\begin{table*}[h]
\caption{\vspace{-0.3em} Segmentation performances on live surgery dataset}
\vspace{-1.5em}
%of our cross-domain strategy and the mainstream strategy
\begin{center}
\begin{tabular}{c||c|c|c|c|c|c}
\hline
Image-to-image & Segmentation & Segmentation & \multicolumn{4}{c}{Segmentation Peformance (mDSC($\%$) / mIoU($\%$))}\\ \cline{4-7}
Translation Model & Train Set & Test Set & DeepLabV3+ \cite{chen2018encoder} & TernausNet-16 \cite{iglovikov2018ternausnet} & DeepLabV3+ \cite{chen2018encoder} & LWANet \cite{ni2019attention}\\
& & & with ResNet50 & with VGG16 & with MobileNet & with MobileNet\\
\hline
%N/A & \multirow{5}{*}{Cadaver} & \multirow{5}{*}{Live} & 38.7/33.2 & 41.4/34.6 & 45.5/38.4 & 29.4/24.3\\
UNIT \cite{liu2017unsupervised} & \multirow{4}{*}{Cadaver} & \multirow{4}{*}{Live} & 34.4/25.8 & 35.6/26.9 & 34.4/26.0 & 35.1/25.9\\
MUNIT \cite{huang2018multimodal} &&& 22.4/17.8 & 18.9/14.2 & 23.8/18.1 & 20.9/15.3\\
CycleGAN \cite{zhu2017unpaired} &&& 62.6/51.4 & 59.9/48.7 & 59.6/48.2 & 57.7/46.2\\
LC-GAN (ours) &&& \textbf{79.9/73.1} & \textbf{75.1/68.1} & \textbf{79.0/71.1} & \textbf{72.8/64.1}\\
\hline
\hline
N/A & Live & Live & 82.7/75.5 & 82.4/75.7 & 83.0/75.7 & 82.1/74.6\\
\hline
\multicolumn{7}{l}{* All segmentation methods were implemented with MAFA \cite{2020arXiv200210675Q}. The first four rows show the results of the cross-domain strategy, }\\
\multicolumn{7}{l}{and the last row shows the results of the mainstream strategy. The bold font indicates the best performance of the cross-domain }\\
\multicolumn{7}{l}{strategy in each column.}
\end{tabular}
\end{center}
\vspace{-2.em}
\label{Tab_seg}
\end{table*}
%The first 4 rows correspond to results from our image-to-image translation-based instrument segmentation strategy. The last row is the mainstream strategy that directly trains and tests segmentation models on the target dataset.
%\multirow{2}{*}{Train Set}

% ablation study
\begin{table}[h]
\caption{\vspace{-0.3em} Ablation studies of LC-GAN with DeepLabV3+(ResNet50)}
\vspace{-1.5em}
%of our cross-domain strategy and the mainstream strategy
\begin{center}
\begin{tabular}{c|c|c|c}
% \begin{tabular}{p{0.08\textwidth}|p{0.08\textwidth}|p{0.08\textwidth}|p{0.13\textwidth}|p{0.13\textwidth}|p{0.13\textwidth}|p{0.13\textwidth}}
\hline
$~\mathcal{L}_{SSim}~$ & $~~\mathcal{L}_{seg}~~$ & Trained & Segmentation Peformance \\
 &  & Backbone & (mDSC($\%$) / mIoU($\%$))\\
\hline
$\times$ & $\times$ & $\times$ & 24.1/16.8\\
$\surd$ & $\times$ & $\times$ & 78.2/71.4\\
$\times$ & $\surd$ & $\times$ & 77.7/70.9\\
$\times$ & $\times$ & $\surd$ & 78.2/71.0\\
$\surd$ & $\surd$ & $\times$ & 79.0/72.2\\
$\surd$ & $\times$ & $\surd$ & 78.4/71.6\\
$\times$ & $\surd$ & $\surd$ & 79.1/72.5\\
$\surd$ & $\surd$ & $\surd$ & \textbf{79.9/73.1}\\
\hline
\multicolumn{4}{l}{* The bold font indicates the best performance in the column. }
% \multicolumn{7}{l}{* The performance of different combinations of three contributions of LC-GAN: 1) structural similarity loss $\mathcal{L}_{SSim}$; 2) segmentation consistency loss $\mathcal{L}_{seg}$; 3) All segmentation methods were implemented with MAFA \cite{2020arXiv200210675Q}. }\\
% \multicolumn{7}{l}{The bold font indicates the best performance of the cross-domain }\\
% \multicolumn{7}{l}{strategy in each column.}
\end{tabular}
\end{center}
\vspace{-2.em}
\label{Tab_ablation}
\end{table}

\textbf{Quantitative evaluation} We used the instrument segmentation performances on the fake-cadaveric surgery images translated from the real-live surgery images for quantitative evaluations. The segmentation models were trained on the cadaveric dataset. 
%We applied four segmentation models (with MAFA \cite{2020arXiv200210675Q}) include DeepLabV3+ \cite{chen2018encoder}, TernausNet \cite{iglovikov2018ternausnet}, MFF+SPP \cite{islam2019real} and LWANet \cite{ni2019attention}.
We applied three segmentation models including DeepLabV3+ \cite{chen2018encoder}, TernausNet \cite{iglovikov2018ternausnet} and LWANet \cite{ni2019attention} with different pre-trained backbone feature extractors and a Multi-angle Feature Aggregation (MAFA) strategy \cite{2020arXiv200210675Q}. The implementation details of the segmentation models are provided in \cite{2020arXiv200210675Q}.
%Instrument segmentation is implemented using four deep CNN models including DeepLabV3+ \cite{chen2018encoder}, TernausNet \cite{iglovikov2018ternausnet}, MFF+SPP \cite{islam2019real} and LWANet \cite{ni2019attention}, and a Multi-angle Feature Aggregation (MAFA) strategy \cite{2020arXiv200210675Q}.
As a comparison, we also performed the mainstream strategy that trains and tests these segmentation models directly with the labeled live surgery dataset. %We studied how close can our strategy (without using the labels of the live surgery dataset) be to these performance upper bounds. 
We used the Dice similarity coefficient (DSC) and Intersection over Union (IoU) to evaluate the instrument segmentation performance \cite{taha2015metrics}
$$
DSC = \frac{2|X\cap Y|}{|X|+|Y|},IoU = \frac{|X\cap Y|}{|X\cup Y|}
$$
where $X$ and $Y$ are the predicted and ground truth instrument segmentations, respectively. 

Table \ref{Tab_seg} shows the segmentation results. We found that segmentation on fake-cadaveric surgery images translated by LC-GAN were better than the results obtained using other compared image-to-image translation models. Our method achieved 15.1$\%$$\sim$19.4$\%$ better mDSC and 17.9$\%$$\sim$22.9$\%$ better mIoU than using CycleGAN. 
Compared with the mainstream strategy that uses the labeled live dataset, our method achieved 2.8$\%$$\sim$9.3$\%$ lower mDSC and 2.4$\%$$\sim$10.5$\%$ lower mIoU. 
%We will discuss potential methods to mitigate the gaps in Section \ref{discussion}.
%The current gaps between our strategy and mainstream strategies are acceptable and show that our method can generate images similar to the real images with a certain guarantee on semantic consistency.

To evaluate the effectiveness of the proposed modules i) generator architectures with trained backbone, ii) structural similarity loss $\mathcal{L}_{SSim}$ and iii) segmentation consistency loss $\mathcal{L}_{seg}$ in LC-GAN, we conducted ablation studies with different configurations. Table \ref{Tab_ablation} shows the evaluation results with DeepLabV3+ \cite{chen2018encoder}. 
% with Resnet-50 \cite{he2016identity} as the backbone
All experiments were implemented with the same LC-GAN hyper-parameters described in Section \ref{implement}. The network failed to converge without any proposed module in limited training epochs, while all three proposed modules helped the network converge faster and achieve acceptable performances. Moreover, combinations of two or three proposed modules led to 0.2$\%$$\sim$2.2$\%$ better mDSC and mIoU than using only one proposed module.

\section{Discussion}\label{discussion}
%Adding the $\mathcal{L}_{SSim}$ slows down the training in each iteration, but make the model converge with less iterations than the original CycleGAN.

% summarize
We propose an image-to-image translation model LC-GAN that achieves better semantic consistency using constraints that encourage structural similarity. %Our method reduces the need to manually label new large datasets for a cross-domain segmentation task by translating a labeled dataset to a relevant but different domain.
Our method reduces the need to manually label new large datasets for a cross-domain segmentation task. 
The training of the proposed image-to-image translation model only requires an unpaired dataset, which can be easily extracted from the surgery videos.

% quantitatively analyze the translation results
Fig. \ref{trans_rst_ex} shows that LC-GAN surpasses other comparison methods to provide the best image-to-image translation results. In contrast, CycleGAN tends to increase instrument sizes in the fake-cadaveric surgery images. This can be explained by the fact that in our dataset the instruments in the cadaveric domain are generally larger than the instruments in the live domain. UNIT and MUNIT fail to capture the correct mapping between the cadaveric and live surgery images. 
%The scenes in the two domains have many differences: there are more types of tools and more complex backgrounds including bone, carbonized tissue and Titanium mesh in the live surgery images. This fact may explain that UNIT and MUNIT cannot discover the correct domain correspondence because they add too strict assumptions to the dataset. 
UNIT and MUNIT are built based on a shared-latent space assumption, i.e. each pair of corresponding images from the two domains can be mapped to a shared-latent space. However, this assumption may be too strict for our dataset because the scenes in the two domains have many differences in both the instrument types and backgrounds.
%
% achieve high quality when train CycleGAN with DSSIM at 40 epoch

Compared with the mainstream strategy, the cross-domain strategy achieved the lower performance. This result is as expected and the current gaps are acceptable. %especially when DeepLabV3+ \cite{chen2018encoder} is used for segmentation.
% This result is consistent with intuition. 
The live and cadaveric datasets have different types of instruments and lightning conditions, so the distribution of the fake-cadaveric images is still different from the real-cadaveric images. Also, although we propose two loss functions to improve semantic consistency between the real images and their corresponding fake images, the semantic consistency has not been fully guaranteed. %The potential strategies to fill this gap in the future will be discussed in Section \ref{discussion}. 
To mitigate the gap, we plan to use a small set of fake-cadaveric images with labels to fine-tune the deep segmentation models. 
% explain our dataset is difficult
%Moreover, it should be noted that the proposed live surgery dataset is very challenging. 
Moreover, the proposed live surgery dataset introduces multiple challenges. When we labeled the dataset, we found that it is much easier to decide the instrument locations by referring to neighboring frames. This points out a future work direction, by estimating the motion flow between the neighboring frames in the video, we can potentially obtain clues of semantic consistency for image-to-image translation.

\section{Conclusions}

In this work, we propose an image-to-image translation model LC-GAN to learn the mapping between two different but relevant image domains. We introduce structural similarity loss and segmentation consistency loss for LC-GAN to improve the semantic consistency during translation. We demonstrate the proposed model in a sinus surgery dataset of cadaveric and live surgery images. Our results show that the proposed method can potentially reduce the need to label more data for surgical instrument segmentation. These results have major implications on the ability to automatically segment and track surgical instruments, leading to improved analysis of surgery as well as enhancement in surgical training.

% \addtolength{\textheight}{-12cm}   % This command serves to balance the column lengths
                                  % on the last page of the document manually. It shortens
                                  % the textheight of the last page by a suitable amount.
                                  % This command does not take effect until the next page
                                  % so it should come on the page before the last. Make
                                  % sure that you do not shorten the textheight too much.

%%%%%%%%%%%%%%%%%%%%%%%%%%%%%%%%%%%%%%%%%%%%%%%%%%%%%%%%%%%%%%%%%%%%%%%%%%%%%%%%
% \section*{APPENDIX}

% Appendixes should appear before the acknowledgment.

% \section*{ACKNOWLEDGMENT}

%%%%%%%%%%%%%%%%%%%%%%%%%%%%%%%%%%%%%%%%%%%%%%%%%%%%%%%%%%%%%%%%%%%%%%%%%%%%%%%%
\bibliographystyle{IEEEtran}
\bibliography{IEEEabrv,IEEEexample}

\end{document}